\documentclass[12pt]{article}
\usepackage{amsfonts}
\usepackage{bm}
\begin{document}

\title{The averaging of weakly nonlocal Symplectic Structures.}

\author{A.Ya. Maltsev}

\date{
\centerline{L.D.Landau Institute for Theoretical Physics,}
\centerline{119334 ul. Kosygina 2, Moscow, maltsev@itp.ac.ru}}

\maketitle

\begin{abstract}
 We consider the averaging of the weakly nonlocal Symplectic 
Structures corresponding to local evolution PDE's in the Whitham 
method. The averaging procedure gives the weakly nonlocal
Symplectic Structure of Hydrodynamic Type for the corresponding
Whitham system. The procedure gives also the "action variables"
corresponding to the wave numbers of $m$-phase solutions of
initial system which give the additional conservation laws
for the Whitham system.
\end{abstract}

\vspace{0.5cm}

 We consider the general weakly nonlocal Symplectic Structures ([1]) 
having the form

\begin{equation}
\label{SympStr}
\Omega_{ij}(x,y)  =  \sum_{k \geq 0} 
\omega^{(k)}_{ij} (\varphi, \varphi_{x}, \dots)  \delta^{(k)}(x-y)  +  
\sum_{s=1}^{g} e_{s} {\delta H^{(s)} \over \delta \varphi^{i}(x)}
\, \nu(x-y) \, {\delta H^{(s)} \over \delta \varphi^{j}(y)}
\end{equation}
where $i,j = 1, \dots , n$, $\nu(x-y) = (1/2) sgn (x-y)$, 
$\delta^{(k)}(x-y) = \partial^{k}/\partial x^{k} \delta (x-y)$, 
$e_{s} = \pm 1$, 
$H^{(s)}[\varphi] = \int_{-\infty}^{+\infty} 
h^{(s)} (\varphi, \varphi_{x}, \dots) dx$ and $\Omega_{ij}(x,y)$ gives 
the closed 2-form on the space of functions 
$\varphi(x) = (\varphi^{1}(x), \dots , \varphi^{n}(x))$. 
We assume also that every term in (\ref{SympStr}) depends on the finite 
number of derivatives $\varphi, \varphi_{x}, \dots$. 
\footnote{Easy to see that the nonlocal part of (\ref{SympStr})
is connected actually with the quadratic form on the space
of functionals $H^{(s)}$ which is written here in the diagonal 
form.}

\vspace{0.5cm}

 We say that the evolution system

\begin{equation}
\label{insyst}
\varphi^{i}_{t}  \,\,\, = \,\,\, Q^{i} (\varphi, \varphi_{x}, \dots)
\end{equation}
admits the Symplectic Structure (\ref{SympStr}) with the Hamiltonian 
function
$H = \int_{-\infty}^{+\infty} h (\varphi, \varphi_{x}, \dots) dx$ 
if the following relation is true

$$\sum_{k \geq 0} \omega^{(k)}_{ij} (\varphi, \varphi_{x}, \dots)
{\partial^{k} \over \partial x^{k}} Q^{j} (\varphi, \varphi_{x}, \dots) +
\sum_{s=1}^{g} e_{s} {\delta H^{(s)} \over \delta \varphi^{i}(x)} D^{-1}
{\delta H^{(s)} \over \delta \varphi^{j}(x)}
Q^{j} (\varphi, \varphi_{x}, \dots) \equiv 
{\delta H \over \delta \varphi^{i}(x)} $$
where $D^{-1}$ is the integration operator defined in the skew-symmetric
way, i.e. $D^{-1} \xi (x) = (1/2) \int_{-\infty}^{x} \xi (y) dy -
(1/2) \int_{x}^{+\infty} \xi (y) dy$. The functionals $H^{(s)}[\varphi]$
should give the conservation laws for the system (\ref{insyst}) in this 
case such that $h^{(s)}_{t} \equiv \partial_{x} J^{(s)}$ for some
$J^{(s)}(\varphi, \varphi_{x}, \dots)$.

\vspace{0.5cm}

 We are going to consider the well-known Whitham averaging method 
[2,3,4,5,6] 
for system (\ref{insyst}) which gives the nonlinear system of equations 
on the slow-modulated parameters $U^{\nu}$ of $m$-phase solutions of 
(\ref{insyst})

\begin{equation}
\label{qpersol}
\varphi^{i}(x,t) \,\,\, = \,\,\,
\Phi^{i} (\omega(U) t + k(U) x + \theta_{0}, U^{1}, \dots, U^{N})
\end{equation}
having the form

\begin{equation}
\label{whithsyst}
U^{\nu}_{T} \,\,\, = \,\,\, V^{\nu}_{\mu}(U) \,\, U^{\mu}_{X} 
\,\,\, , \,\,\,\,\, 
\nu, \mu = 1, \dots, N
\end{equation}
(Whitham system). The functions $\Phi^{i}(\theta,U)$ here are 
$2\pi$-periodic functions with respect to the variables 
$(\theta^{1}, \dots, \theta^{m}) = \theta$ depending on the additional 
parameters $U^{1}, \dots, U^{N}$. The ($m$-component) vectors $\omega(U)$ 
and $k(U)$ play the role of frequences and "wave numbers" for the 
quasiperiodic solutions (\ref{qpersol}). In Whitham method 
the parameters $U^{\nu}$ become the functions of the
"slow variables" $X = \epsilon x$, $T = \epsilon t$, 
$\epsilon \rightarrow 0$ and the system (\ref{whithsyst}) gives the 
necessary conditions for the construction of corresponding asymptotic
solution of (\ref{insyst}).

\vspace{0.5cm}

 We will give here the procedure of "averaging"of the Symplectic Structure
(\ref{SympStr}) giving the weakly nonlocal Symplectic Structure of
Hydrodynamic Type for system (\ref{whithsyst}).

\vspace{0.5cm}

{\bf Definition 1.} 
{\it We call the weakly nonlocal Symplectic Structure of Hydrodynamic 
Type the weakly nonlocal Symplectic Structure having the form

\begin{equation}
\label{HTstr}
\Omega_{\nu\mu}(X,Y) = \sum_{s,p=1}^{M} \kappa_{sp} 
{\partial f^{(s)} \over \partial U^{\nu}}(X) \,\, \nu (X-Y) \,\,
{\partial f^{(p)} \over \partial U^{\mu}}(Y)
\end{equation} 
where $f^{(s)}(U)$ are some functions of variables $U^{1}, \dots, U^{N}$
and $\kappa_{sk}$ is some constant symmetric bilinear form
($\nu,\mu = 1, \dots, N$).
}

\vspace{0.5cm}

 Let us formulate now the procedure giving the Symplectic Structure 
(\ref{HTstr}) for the Whitham system. We introduce first the "averaging"
of any local function $f (\varphi, \varphi_{x}, \dots)$ over the 
"invariant tori" corresponding to the family (\ref{qpersol}). Namely

$$\langle f \rangle (U) = \int_{0}^{2\pi} \dots \int_{0}^{2\pi}
f \left(\, \Phi(\theta, U), \, 
k^{\alpha}(U) \, \Phi_{\theta^{\alpha}}(\theta, U),
\dots \right) \,\, {d^{m} \theta \over (2\pi)^{m}} $$

 Let us introduce the "extended functional space" of functions 
$\varphi (\theta, x)$ $2\pi$-periodic w.r.t. to each $\theta^{\alpha}$
at every fixed $x$ and
define the "extended" Symplectic Form 
${\tilde \Omega}_{ij}(\theta, \theta^{\prime}, x, y) =
\delta (\theta - \theta^{\prime}) \Omega_{ij}(x, y)$.

\vspace{0.5cm}

 We introduce also the special functions $T^{(s)}_{\alpha}$ connected
with the nonlocal part of the Symplectic Structure 
${\tilde \Omega}_{ij}(\theta, \theta^{\prime}, x, y)$

\begin{equation}
\label{tsa}
T^{(s)}_{\alpha}(\varphi, \varphi_{x}, \varphi_{\theta^{\alpha}}, \dots)
= \sum_{k \geq 1} \sum_{p=0}^{k-1} (-1)^{p} \left(
{\partial h^{(s)} \over \partial \varphi^{i}_{kx}}\right)_{px}
\varphi_{\theta^{\alpha},(k-p-1)x} 
\end{equation}
(here $f_{kx} \equiv \partial^{k} f/\partial x^{k}$ and we assume
summation over the repeated indices).

\vspace{0.5cm}

{\bf Lemma 1.}

{\it 1) For any Symplectic Form (\ref{SympStr}) we have the relations

$$\varphi^{i}_{\theta^{\alpha}} \sum_{k \geq 0} 
\omega^{(k)}_{ij} (\varphi, \varphi_{x}, \dots) 
\varphi^{j}_{\theta^{\beta},kx} + \sum_{s=1}^{g} e_{s} \left(
h^{(s)}_{\theta^{\beta}} T^{(s)}_{\alpha} - 
h^{(s)}_{\theta^{\alpha}} T^{(s)}_{\beta} + 
\left(T^{(s)}_{\alpha}\right)_{x} T^{(s)}_{\beta} \right) \equiv $$

$$\equiv {\partial \over \partial \theta^{\gamma}}
Q^{\gamma}_{\alpha\beta} (\varphi, \dots) + {\partial \over \partial x}
A_{\alpha\beta} (\varphi, \dots) $$
for some functions $Q^{\gamma}_{\alpha\beta} (\varphi, \dots)$,
$A_{\alpha\beta} (\varphi, \dots)$

 2) The functions $A_{\alpha\beta} (\varphi, \dots)$ (defined modulo
constant functions) can be normalized in such a way that
$A_{\alpha\beta} (\varphi, \dots) \equiv 0$ for any 
$\varphi(\theta, x)$ depending on $x$ only (and constant with respect
to $\theta$ at every fixed $x$).
}

\vspace{0.5cm}

 We will assume now that we have $m$ linearly independent commuting flows

$$\varphi^{i}_{t^{\alpha}} 
\,\, \, = \,\,\, Q^{i}_{\alpha} (\varphi, \varphi_{x}, \dots)\,\, ,
\,\,\,\,\, \alpha = 1, \dots, m $$ 
for the system (\ref{insyst}) which leave the
family (\ref{qpersol}) invariant generating the linear shifts of 
initial phases
$\theta_{0}^{\alpha}$ on it with the frequences $\omega_{(\alpha)}(U)$.
We require also that the functionals $H^{(s)} [\varphi]$ give the 
conservation laws for the flows $\varphi^{i}_{t^{\alpha}}$ such that

\begin{equation}
\label{hja}
h^{(s)}_{t^{\alpha}} \,\,\, \equiv \,\,\, 
\partial_{x} J^{(s)}_{\alpha}
\end{equation}
for some
local functions $J^{(s)}_{\alpha} (\varphi, \varphi_{x}, \dots)$.

 Besides that, we will require that the matrix 
$||\omega_{(\alpha)}^{\beta}(U)||$ is nondegenerate and has the 
inverse matrix $||\gamma_{\alpha}^{\beta}(U)||$ such that

\begin{equation}
\label{gammamatr}
\gamma^{\delta}_{\alpha}(U) \,\, \omega^{\beta}_{(\delta)}(U) 
\,\,\, \equiv \,\,\,  \delta^{\beta}_{\alpha}
\end{equation}

\vspace{0.5cm}

{\bf Theorem 1.}

{\it
Under the conditions formulated above the Whitham system 
(\ref{whithsyst}) has the weakly nonlocal Symplectic Structure of
Hydrodynamic Type

$$\Omega_{\nu\mu}(X,Y) = \sum_{\alpha = 1}^{m} \left(
{\partial k^{\alpha} \over \partial U^{\nu}}(X) \,\, \nu (X-Y) \,\,
{\partial I_{\alpha} \over \partial U^{\mu}}(Y) +
{\partial I_{\alpha} \over \partial U^{\nu}}(X) \,\, \nu (X-Y) \,\,
{\partial k^{\alpha} \over \partial U^{\mu}}(Y) \right) + $$

\begin{equation}
\label{av2form}
+ \sum_{s=1}^{g} e_{s} 
{\partial \langle h^{(s)} \rangle \over \partial U^{\nu}}(X) 
\,\, \nu (X-Y) \,\,
{\partial \langle h^{(s)} \rangle \over \partial U^{\mu}}(Y) 
\end{equation}
with Hamiltonian function 
$H = \int_{-\infty}^{+\infty} \langle h \rangle dX$ where the
"action variables" $I_{\alpha}(U)$ are given through the formulas:

$${\partial I_{\alpha} \over \partial U^{\nu}} \,\,\, = \,\,\,
 {\partial k^{\beta} \over \partial U^{\nu}} \left[ 
- \langle A_{\alpha\beta} \rangle + \gamma^{\delta}_{\alpha} 
\sum_{s=1}^{g} e_{s} 
\left( \langle T^{(s)}_{\beta} J^{(s)}_{\delta} \rangle -
\langle T^{(s)}_{\beta} \rangle \langle J^{(s)}_{\delta} \rangle 
\right) \right. \,\, - \hspace{5cm}$$

\vspace{-0.5cm}

$$ \hspace{7cm} \left. - \,\, {1 \over 2} 
\gamma^{\delta}_{\alpha} \gamma^{\zeta}_{\beta} \sum_{s=1}^{g} e_{s}
\left( \langle J^{(s)}_{\delta} J^{(s)}_{\zeta} \rangle -
\langle J^{(s)}_{\delta} \rangle \langle J^{(s)}_{\zeta} \rangle 
\right) \right] + $$

$$+ \langle \Phi^{i}_{U^{\nu}} \sum_{k \geq 0} 
\omega^{(k)}_{ij} (\varphi, \dots) \varphi^{j}_{\theta^{\alpha},kx}
\rangle - \sum_{s=1}^{g} e_{s} \langle \Phi^{i}_{U^{\nu}}
{\delta H^{(s)} \over \delta \varphi^{i}(x)} T^{(s)}_{\alpha}
\rangle + $$

\begin{equation}
\label{actvar}
+ \sum_{s=1}^{g} e_{s} \gamma^{\beta}_{\alpha} \left( \langle
\Phi^{i}_{U^{\nu}} {\delta H^{(s)} \over \delta \varphi^{i}(x)}
J^{(s)}_{\beta} \rangle - \langle
\Phi^{i}_{U^{\nu}} {\delta H^{(s)} \over \delta \varphi^{i}(x)}
\rangle \langle J^{(s)}_{\beta} \rangle \right) 
\end{equation}
and the functions $A_{\alpha\beta}$ are normalized according to 
Lemma 1.
}

\vspace{0.5cm}

{\bf Definition 2.}
{\it We call the form (\ref{av2form}) the averaging of the 
weakly nonlocal Symplectic Structure (\ref{SympStr}) 
on the family of $m$-phase solutions of system (\ref{insyst}).
}

\vspace{0.5cm}

 Let us add also that the functionals 
$\int_{-\infty}^{+\infty} I_{\alpha} dX$ as well as
$\int_{-\infty}^{+\infty} k^{\alpha} dX$, 
$\int_{-\infty}^{+\infty} \langle h^{(s)} \rangle dX$ and
$H = \int_{-\infty}^{+\infty} \langle h \rangle dX$ give the
conservation laws for the Whitham system (\ref{whithsyst}).

\vspace{1.5cm}

 We will give now another variant of the averaging procedure
of the form (\ref{SympStr}) using the averaging of weakly
nonlocal 1-forms on the space $\varphi(x)$.

\vspace{0.5cm}

 Let us consider the 1-forms
$\omega_{i}[\varphi](x)$ on the space of functions
$\varphi^{i}(x)$, $i = 1, \dots, n$ having the form

\begin{equation}
\label{wnonl1form}
\omega_{i}[\varphi](x) \, = \,
c_{i} (\varphi,\varphi_{x}, \dots) \, - \,
{1 \over 2} \, \sum_{s=1}^{g} e_{s} \,
{\delta H^{(s)} \over \delta \varphi^{i}(x)} \,
\int_{-\infty}^{+\infty} \nu (x-y) \,
h^{(s)} (\varphi,\varphi_{y}, \dots) \, dy
\end{equation}
where $H^{(s)}[\varphi] \, = \, \int_{-\infty}^{+\infty}
h^{(s)} (\varphi,\varphi_{x}, \dots) \, dx$.

 The action of the forms $\omega_{i}[\varphi](x)$
on the "tangent vectors" $\xi^{i}[\varphi](x)$ is defined
in the natural way

$$(\bm{\omega}, \bm{\xi}) [\varphi]\, = \,
\int_{-\infty}^{+\infty} \omega_{i}[\varphi](x) \,
\xi^{i}[\varphi](x) \, dx $$
 
 The forms (\ref{wnonl1form}) are closely connected with the 
weakly nonlocal 2-forms (\ref{SympStr}). Namely, let us 
consider the external derivative of the form
$\omega_{i}[\varphi](x)$:

$$\left[ d \bm{\omega} \right]_{ij} (x,y) \, = \,
{\delta \omega_{j}[\varphi](y) \over \delta \varphi^{i}(x)}  
\,\, - \,\,
{\delta \omega_{i}[\varphi](x) \over \delta \varphi^{j}(y)} $$

\vspace{0.5cm}

{\bf Lemma 2.}

{\it The external derivative
$\left[ d \bm{\omega} \right]_{ij} (x,y)$ is the closed 2-form
having the form (\ref{SympStr}) with some local functions
$\omega_{ij}^{(k)} (\varphi,\varphi_{x}, \dots)$.
}

\vspace{0.5cm}

 Let us formulate also the opposite statement

\vspace{0.5cm}

{\bf Lemma 3.}

{\it Every closed 2-form (\ref{SympStr}) can be locally
represented as the external derivative of some 1-form
(\ref{wnonl1form}) on the space $\varphi(x)$.
}

\vspace{0.5cm}

 We are going to give now the procedure of averaging of
1-forms (\ref{wnonl1form}) connected with the averaging of
the Symplectic structures (\ref{SympStr}). Namely,
we will assume now that the form $\Omega_{ij}(x,y)$ is
represented as the external derivative of the form
(\ref{wnonl1form}). The corresponding procedure of averaging
of the form (\ref{wnonl1form}) should then give the weakly
nonlocal 1-form of "Hydrodynamic type" which is connected
with the form $\Omega_{\nu\mu} (X,Y)$ in the same way.

\vspace{0.5cm}

{\bf Definition 3.} {\it We call the form
$\omega_{\nu}[{\bf U}](X)$ on the space of functions
$U^{1}(X), \dots, U^{N}(X)$ the weakly nonlocal 1-form of
Hydrodynamic type if it has the form

$$\omega_{\nu}[{\bf U}](X) \,\, = \,\, - {1 \over 2}
\sum_{s,p=1}^{M} \kappa_{sp} \,
{\partial f^{(s)} \over \partial U^{\nu}} ({\bf U}(X)) \,
\int_{-\infty}^{+\infty} \nu(X-Y) \, f^{(p)}({\bf U}(Y)) \,
dY$$
for some functions $f^{(s)}({\bf U})$ and the quadratic form
$\kappa_{sp}$.
}

\vspace{0.5cm}

{\bf Definition 4.} {\it We call the 1-form

$$\omega_{\nu}(X) \,\, = \,\, - \,
{\partial k^{\alpha} \over \partial U^{\nu}}(X) \,
\int_{-\infty}^{+\infty} \nu(X-Y) \, I_{\alpha}(Y) \, dY
\,\, - $$

\begin{equation}
\label{av1form}
- \,\, {1 \over 2} \, \sum_{s=1}^{g} e_{s} \,
{\partial \langle h^{(s)} \rangle \over \partial U^{\nu}}(X)
\, \int_{-\infty}^{+\infty} \nu(X-Y) \,
\langle h^{(s)} \rangle (Y) \, dY
\end{equation}
where $I_{\alpha}({\bf U})$ are defined by the formulas

\begin{equation}
\label{formactvar}
I_{\alpha}({\bf U}) \,\, = \,\,
\langle c_{i} \, \varphi^{i}_{\theta^{\alpha}} \rangle \,
+ \, {1 \over 2} \, \gamma^{\delta}_{\alpha}({\bf U}) \, 
\sum_{s=1}^{g} e_{s} \,
\left[ \langle h^{(s)} \, J^{(s)}_{\delta} \rangle \, - \,
\langle h^{(s)} \rangle \langle J^{(s)}_{\delta} \rangle
\right] \, - \, {1 \over 2} \, \sum_{s=1}^{g} e_{s} \,
\langle h^{(s)} \, T^{(s)}_{\alpha} \rangle
\end{equation}
- the averaging of the 1-form
(\ref{wnonl1form}) on the family of $m$-phase solutions of
(\ref{insyst}).
}

\vspace{0.5cm}

{\bf Theorem 2.}

{\it The averaged forms $\Omega_{\nu\mu}(X,Y)$ and
$\omega_{\nu}(X)$ are connected by the relation

$$\Omega_{\nu\mu}(X,Y) \,\,\, = \,\,\,
[d \, \bm{\omega} ]_{\nu\mu} (X,Y) \,\,\, = \,\,\,
{\delta \omega_{\mu}(Y) \over \delta U^{\nu}(X)} 
\,\, - \,\, 
{\delta \omega_{\nu}(X) \over \delta U^{\mu}(Y)} $$
}

\vspace{0.5cm}

 The quantities $I_{\alpha}({\bf U})$ defined in 
(\ref{formactvar}) coincide with the same quantities defined in
(\ref{actvar}).

\vspace{0.5cm}

 It's not difficult to see also that the form (\ref{av1form})
differs from the form

$$\omega^{\prime}_{\nu}(X)  =  -  {1 \over 2} \, 
{\partial k^{\alpha} \over \partial U^{\nu}}(X) 
\int_{-\infty}^{+\infty} \! \nu(X-Y) \, I_{\alpha}(Y) \, dY
\, - \, {1 \over 2} \,
{\partial I_{\alpha} \over \partial U^{\nu}}(X) 
\int_{-\infty}^{+\infty} \! 
\nu(X-Y) \, k^{\alpha}(Y) \, dY \, -$$

$$- \,\, {1 \over 2} \, \sum_{s=1}^{g} e_{s} \,
{\partial \langle h^{(s)} \rangle \over \partial U^{\nu}}(X)
\, \int_{-\infty}^{+\infty} \nu(X-Y) \,
\langle h^{(s)} \rangle (Y) \, dY$$
just by exact 1-form

$$\bm{\omega} \,\, - \,\, \bm{\omega}^{\prime} \,\,\, = \,\,\,
d \,\,\, {1 \over 2} \, \int_{-\infty}^{+\infty}
\int_{-\infty}^{+\infty} I_{\alpha}(X) \, \nu(X-Y) \,
k^{\alpha}(Y) \,\, dX \, dY $$

\vspace{0.5cm}

 The formulas (\ref{av1form}),
(\ref{formactvar}) give another procedure for the
averaging of 2-forms $\Omega_{ij}(x,y)$ represented in the
form of the external derivatives of weakly-nonlocal 1-forms
$\omega_{i}(x)$.

  We can also write the formal (nonlocal) Lagrangian formalism 
for the Whitham equations in the form

$$\delta \,\, \int \left[ \omega_{\nu}(X) \,
U^{\nu}_{T}(X) \, - \, \langle h \rangle ({\bf U}) \right]
\,\, dX \, dT \,\,\, = \,\,\, 0 $$

\vspace{0.5cm}

  The work was supported by Grant of President of
Russian Federation (MK - 1375.2003.02), Russian Foundation
for Basic Research (grant RFBR 03-01-00368), and Russian
Science Foundation.

\vspace{0.5cm}

\noindent 
[1] A.Ya.Maltsev, S.P. Novikov.
Physica D 156 (2001) 53-80.

\noindent
[2] G. Whitham, Linear and Nonlinear Waves.
Wiley, New York (1974).

\noindent
[3]  Luke J.C., {\it Proc. Roy. Soc. London Ser. A},
{\bf 292}, No. 1430, 403-412 (1966).

\noindent
[4] B.A.Dubrovin, S.P.Novikov.,
{\it Soviet Math. Dokl.}, Vol. 27, (1983) No. 3, 665-669.

\noindent
[5] B.A. Dubrovin, S.P. Novikov.,
{\it Russian Math. Survey}, {\bf 44}:6 (1989), 35-124.

\noindent 
[6] B.A. Dubrovin, S.P. Novikov.,
{\it Sov. Sci. Rev. C, Math. Phys.}, 1993, V.9. part 4.
P. 1-136.

\end{document}